\documentclass{jcgt}

\setciteauthor{Russell A. Brown}
\setcitetitle{A Dynamic, Self-balancing k-d Tree}

\accepted{submitted}{accepted}{published}{Editor Name}{vol}{issue}{1}{1}{year}
\seturl{http://jcgt.org/published/vol/issue/num/}

\usepackage[usenames,dvipsnames]{xcolor}
\usepackage{tcolorbox}
\usepackage{tabularx}
\usepackage{array}
\usepackage{tabto}
\usepackage{colortbl}
\tcbuselibrary{skins}

\newcolumntype{Y}{>{\raggedleft\arraybackslash}X}
\tcbset{tab1/.style={fonttitle=\bfseries\large,fontupper=\normalsize\sffamily,
colback=yellow!10!white,colframe=red!75!black,colbacktitle=Salmon!40!white,
coltitle=black,center title,freelance,frame code={
\foreach \n in {north east,north west,south east,south west}
{\path [fill=red!75!black] (interior.\n) circle (3mm); };},}}

\tcbset{tab2/.style={enhanced,fonttitle=\bfseries,fontupper=\normalsize\sffamily,
colback=white!10!white,colframe=black!50!black,colbacktitle=Salmon!40!white,
coltitle=black,center title}}

\usepackage{listings}

\usepackage{xcolor}

\definecolor{codegreen}{rgb}{0,0.6,0}
\definecolor{codegray}{rgb}{0.5,0.5,0.5}
\definecolor{codepurple}{rgb}{0.58,0,0.82}
\definecolor{backcolour}{rgb}{0.99,0.98,0.98}

\lstdefinestyle{mystyle}{
    backgroundcolor=\color{backcolour},   
    commentstyle=\color{codegreen},
    keywordstyle=\color{magenta},
    numberstyle=\tiny\color{codegray},
    stringstyle=\color{codepurple},
    basicstyle=\ttfamily\footnotesize,
    breakatwhitespace=true,         
    breaklines=true,                 
    captionpos=b,                    
    keepspaces=true,                 
    numbers=left,                    
    numbersep=5pt,                  
    showspaces=false,                
    showstringspaces=false,
    showtabs=false,                  
    tabsize=2
}

\begin{document}

\title{A Dynamic, Self-balancing \emph{k}-d Tree}

\author
       {Russell A. Brown}

\maketitle

\begin{abstract}
\small

The original description of the \emph{k}-d tree recognized that rebalancing techniques, used for building an AVL or red-black tree, are not applicable to a \emph{k}-d tree, because these techniques involve cyclic exchange of tree nodes that violates the invariant of the \emph{k}-d tree. For this reason, a static, balanced \emph{k}-d tree is often built from all of the \emph{k}-dimensional data en masse. However, it is possible to build a dynamic \emph{k}-d tree that self-balances when necessary after insertion or deletion of each \emph{k}-dimensional datum. This article describes insertion, deletion, and rebalancing algorithms for a dynamic, self-balancing \emph{k}-d tree, and measures their performance.

\end{abstract}

\section{Introduction} 
\label{sec:Introduction}

Bentley introduced the \emph{k}-d tree as a binary search tree that stores \emph{k}-dimensional data \cite{Bentley}.  Like a binary search tree, a \emph{k}-d tree partitions a set of data at each recursive level of the tree.  Unlike a binary search tree that uses only one key for all levels of the tree, a \emph{k}-d tree uses $k$ keys and cycles through the keys for successive levels of the tree.  For example, to build a \emph{k}-d tree from three-dimensional tuples that represent $\left(x,y,z\right)$ coordinates, the keys would be cycled as $x,y,z,x,y,z...$  for successive levels of the \emph{k}-d tree. A more elaborate scheme for cycling the keys chooses the coordinate that has the widest dispersion or largest variance to be the key for a particular level of recursion \cite{Friedman} but this article assumes that the keys are cycled as $x,y,z,x,y,z...$ in order to simplify the exposition.

Bentley proposed that the $x$-, $y$-, and $z$-coordinates not be used as keys independently of one another, but instead that $x$, $y$, and $z$ form the most significant portions of the respective super keys $x$:$y$:$z$, $y$:$z$:$x$, and $z$:$x$:$y$ that represent cyclic permutations of $x$, $y$, and $z$.  The symbols for these super keys use a colon to designate the concatenation of the individual $x$, $y$ and $z$ values.  For example, the symbol $z$:$x$:$y$ represents a super key wherein $z$ is the most significant portion of the super key, $x$ is the middle portion of the super key, and $y$ is the least significant portion of the super key.

A dynamic \emph{k}-d tree implementation reported previously \cite{Procopiuc} maintained balance via multiple, static \emph{k}-d trees \cite{Bentley1979}. The present article reports the first implementation of balance maintained via a single \emph{k}-d tree.

\begin{figure}[h]
\centering
\centerline{\includegraphics*[trim = {0.96in, 0.1in, 0.80in, 1.36in}, clip, width=\columnwidth]{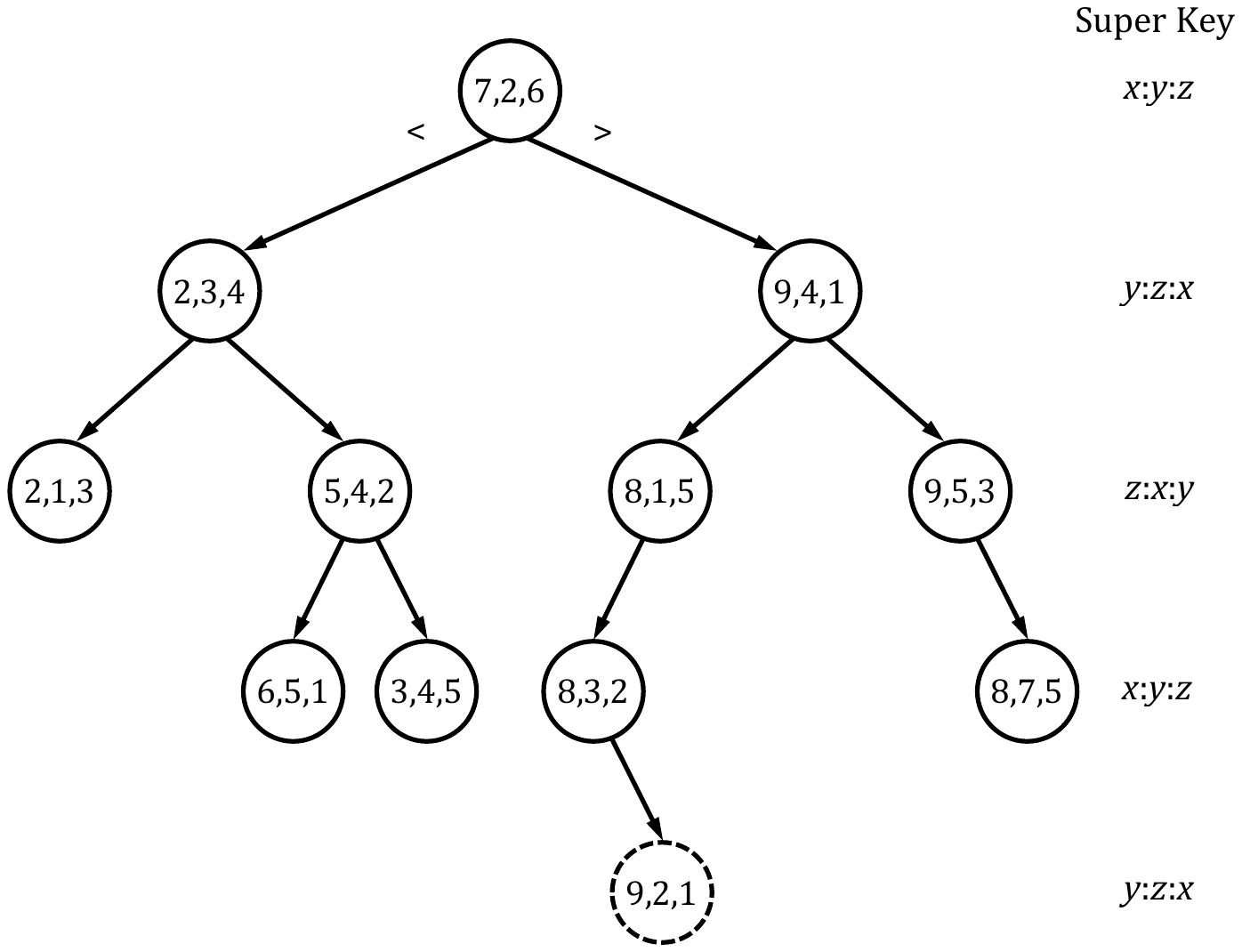}}
\caption{An unbalanced \emph{k}-d tree built by partitioning via $x$:$y$:$z$, $y$:$z$:$x$, and $z$:$x$:$y$ super keys}
\label{fig:unbalanced_tree}
\end{figure}

Figure \ref{fig:unbalanced_tree} depicts a dynamic \emph{k}-d tree that partitions 12 $\left( x,y,z \right)$ tuples via an $x$:$y$:$z$ super key at the root of the tree, then partitions them via $y$:$z$:$x$ super keys at the first level below the root, then partitions them via $z$:$x$:$y$ super keys at the second level below the root, and then partitions them via $x$:$y$:$z$ super keys at the third level below the root. The $<$ and $>$ symbols specify that each left or right child node has a smaller or larger super key respectively than its parent.

This \emph{k}-d tree is not balanced. Node $\left( 9,2,1 \right)$ depicted with a dashed border, and that has been added to the nascent tree as a new leaf node, violates the balance of the tree, as defined below. Hence, the tree must be rebalanced, as discussed subsequently.

\subsection{Two Definitions of Balance}
\label{sec:Balance}

The balance at each node of a dynamic \emph{k}-d tree is computed from the heights of the node's left and right subtrees, where height is defined as the maximum path length from the node to the bottom of the tree. The height of a \lstinline{null} child pointer is 0. The height of a leaf node is 1, and the height of a non-leaf node is 1 plus the greater of the heights of its two children. A node's height is assigned to that node's \lstinline{height} field.

This definition of height permits definition of two types of balance. The first type of balance, AVL balance, requires that each node's left and right subtrees differ in height by at most one node \cite{Adelson}.

The second type of balance, red-black balance, requires that, at each node of the tree, the node's left and right subtrees differ in height by a factor of 2 or less. This requirement derives from the following features of a red-black tree: (1) a node is colored either red or black, (2) neither child of a red node may be red, and (3) every path from any node to the bottom of the tree must contain the same number of black nodes \cite{Guibas}. Given these features, the shortest path from any node to the bottom of a red-black tree contains only black nodes, whereas the longest path contains alternating red and black nodes. Because every path from any node to the bottom of the tree must contain the same number of black nodes, the longest path is twice the length of the shortest path \cite{Drozdek}. For a \emph{k}-d tree, red-black balance doesn't require coloring each node in the \emph{k}-d tree via a \lstinline{color} field. It requires only that each node's left and right subtrees differ in height by a factor of 2 or less.

The algorithm described in the present article may be configured to use either AVL balance or red-black balance, but not both together. One caveat relative to red-black balance is that, for a node that has only one child, the balance of the other, \lstinline{null} child is 0, so the balance of the only child is always greater than twice the balance of the \lstinline{null} child. This problem is resolved by reverting to AVL balance for this case.

\subsection{Insertion into a \emph{k}-d Tree and Subsequent Rebalancing}

Insertion into a \emph{k}-d tree occurs in a similar manner to insertion into a standard binary search tree (i.e., a 1-d binary search tree) \cite{Drozdek} \cite{Weiss}. To insert a \emph{k}-dimensional tuple, the tree is searched recursively for that tuple, comparing the tuple's super key to the tree node's $x$:$y$:$z$, $y$:$z$:$x$, or $z$:$x$:$y$ super key at each level of the tree as the tuple descends the tree. If the search arrives at the bottom of the tree without finding the tuple, the tuple is inserted into the tree as a new leaf node.

\begin{figure}[h]
\centering
\centerline{\includegraphics*[trim = {0.96in, 1.6in, 0.80in, 1.36in}, clip, width=\columnwidth]{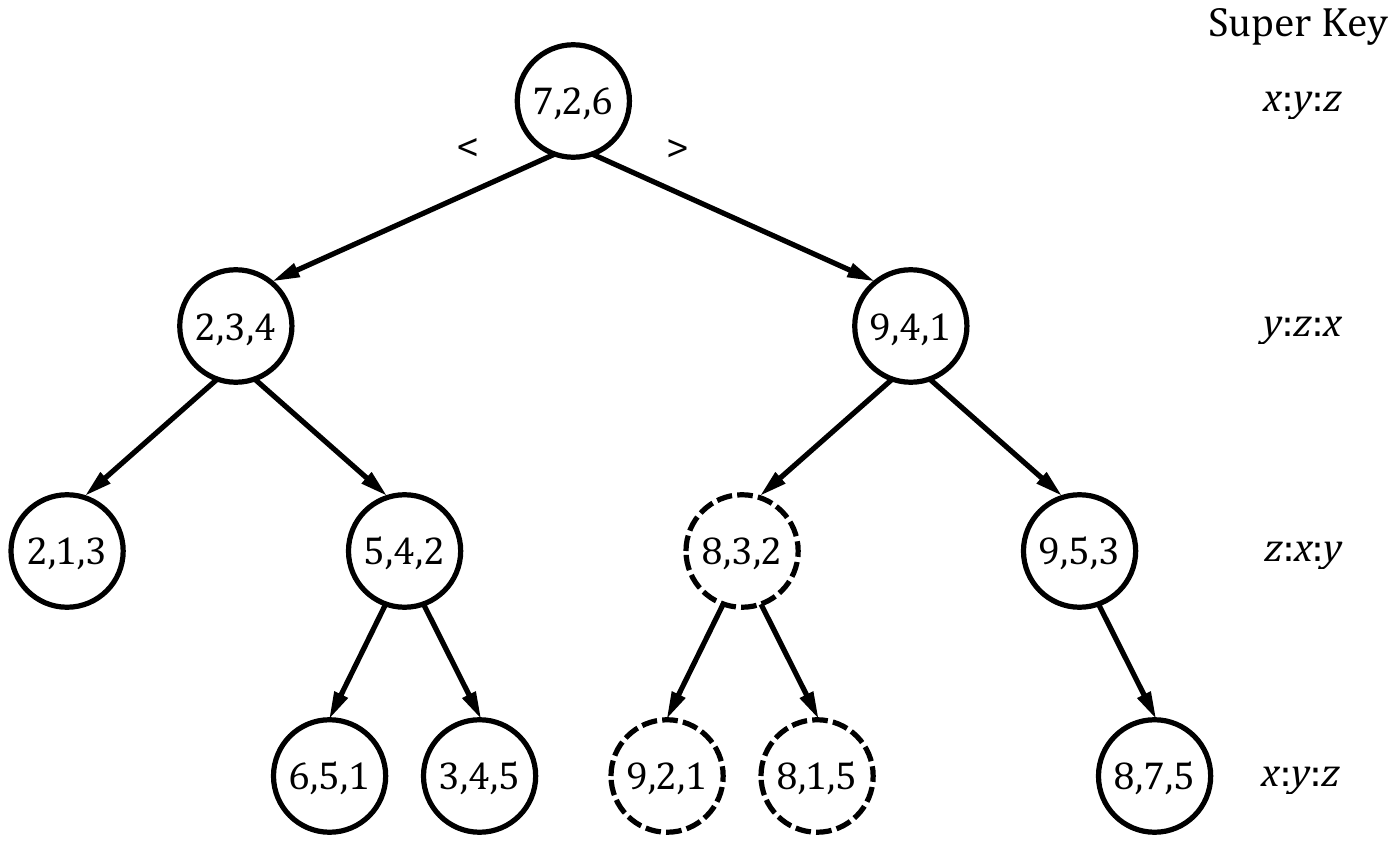}}
\caption{Rebuilding the subtree rooted at node $\left( 8,1,5 \right)$ in Figure \ref{fig:unbalanced_tree} rebalances the \emph{k}-d tree}
\label{fig:balanced_tree}
\end{figure}

Then, as the recursion unwinds, the height is computed at each node along the return path to the root of the tree, and the balance is checked at each node to determine whether rebalancing is required at that node. In Figure \ref{fig:unbalanced_tree}, the difference in child subtree heights is 0, 1, and 2 at nodes $\left( 9,2,1 \right)$; $\left( 8,3,2 \right )$; and $\left( 8,1,5 \right)$ respectively. Hence, the subtree rooted at node $\left( 8,1,5 \right)$ must be rebalanced, as required by the AVL or red-black criterion because both criteria enforce a subtree height difference of at most one node. This subtree is rebalanced by rebuilding it \cite{Overmars} to create the subtree shown in Figure \ref{fig:balanced_tree} that contains nodes $\left( 9,2,1 \right)$; $\left( 8,3,2 \right )$; and $\left( 8,1,5 \right)$ depicted with dashed borders, and whose respective $z$:$x$:$y$ super keys have relative order $1$:$9$:$2$ $<$ $2$:$8$:$3$ $<$ $5$:$8$:$1$. Rebuilding a subtree that contains 3 nodes or fewer, such as this subtree, occurs frequently and requires comparison of at most 3 super keys \cite{Stepanov}. Rebuilding a subtree that contains more than 3 nodes requires a static tree-building algorithm, such as a $O \left[ n \log \left( n \right) \right]$ parallel algorithm described previously \cite{Brown2015}.

\newpage

\subsection{Deletion from a \emph{k}-d Tree and Subsequent Rebalancing}
\label{sec:Deletion}

Deletion from a \emph{k}-d tree begins in a similar manner to, but is more complicated than, deletion from a 1-d binary search tree \cite{Drozdek},  \cite{Rocca} \cite{Weiss}. To delete a \emph{k}-dimensional tuple, the tree is searched recursively for that tuple, comparing the tuple's super key to the node's $x$:$y$:$z$, $y$:$z$:$x$, or $z$:$x$:$y$ super key at each level of the tree. If the search finds the tuple, the tuple is deleted from the tree as one of three cases: (1) a leaf node with no children, (2) a node with one child, and (3) a node with two children. After deletion of the node, the height is computed at each node along the return path to the root of the tree as the recursion unwinds, and the balance is checked at each node to determine whether rebalancing is required at that node.

The first case is deletion of a leaf node that has no children. The leaf node is excised from the tree. Then the recursion unwinds, with computation of the height and potential rebalancing at each node along the return path to the root of the tree.

The second case is deletion of a node that has one child. This case must be treated differently for a \emph{k}-d tree than for a 1-d tree. In a 1-d tree, a 1-child node is replaced by its only child. However, for a \emph{k}-d tree, replacing a 1-child node by its child violates the invariant of the subtree rooted at that child \cite{Drozdek} \cite{Rocca}. A 1-child node, whose super key is $x$:$y$:$z$, has a child whose super key is $y$:$z$:$x$. So moving that child to the next higher level in the tree to replace its parent, as is done in a 1-d tree, would require that the child's super key be $x$:$y$:$z$ instead of $y$:$z$:$x$.

\newpage

This problem is solved by replacing a 1-child node by either its immediate predecessor node or its immediate successor node, similar to the replacement of a node that has two children in a 1-d tree. A 1-child node that has only a less-than child is replaced by its immediate predecessor, whereas a 1-child node that has only a greater-than child is replaced by its immediate successor \cite{Drozdek} \cite{Rocca}.

However, finding the predecessor or successor node is more complicated for a \emph{k}-d tree than for a 1-d tree \cite{Drozdek}. The predecessor node to a particular node in a 1-d tree is found by starting with the less-than child to that node, and recursively following the greater-than child pointer until it is \lstinline{null}, because  the null-pointer node has the largest key in the less-than subtree \cite{Wirth}. The successor node to a particular node is found in a similar manner by starting with the greater-than child to that node, and recursively following the less-than child pointer until it is \lstinline{null}.

\begin{figure}[h]
\centering
\centerline{\includegraphics*[trim = {1.7in, 0.1in, 1.2in, 1.24in}, clip, width=\columnwidth]{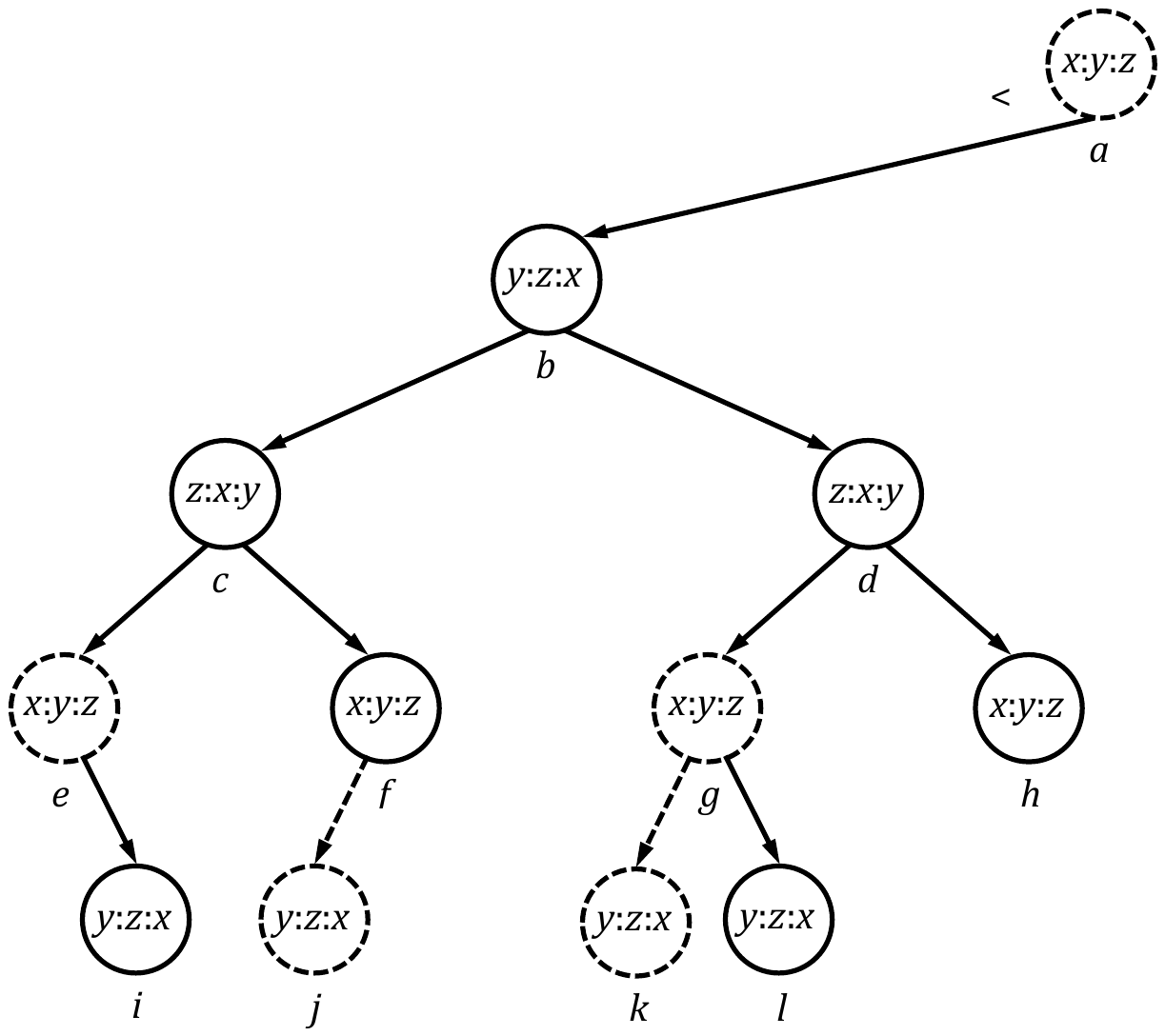}}
\caption{Finding the predecessor node in a \emph{k}-d tree: solid arrows designate child pointers that must be followed recursively, and solid borders designate potential predecessor nodes.}
\label{fig:predecessor}
\end{figure}

In contrast, for a \emph{k}-d tree, recursively following only the greater-than or less-than child pointer does not find the immediate predecessor or successor node respectively. Figure \ref{fig:predecessor} shows the process of finding the immediate predecessor node in a \emph{k}-d tree. 

In Figure \ref{fig:predecessor}, node $a$, which has an $x$:$y$:$z$ super key, is a 1-child node to be replaced by its immediate predecessor node. The search for that predecessor node begins with node $b$, which is node $a$'s less-than child. The subtree rooted at node $b$ must be searched to find the node that has the largest $x$:$y$:$z$ super key, because that node is the immediate predecessor node.

Because node $b$ has a $y$:$z$:$x$ super key, its $x$:$y$:$z$ sorted order relative to node $a$ is unknown, so it must be inspected as a potential predecessor node to node $a$. Also, because of its $y$:$z$:$x$ super key, node $b$ cannot eliminate its children (nodes $c$ and $d$) from recursive search for node $a$'s predecessor node.

Because nodes $c$ and $d$ have $z$:$x$:$y$ super keys, their $x$:$y$:$z$ sorted orders relative to node $a$ are unknown, so they must be inspected as potential predecessor nodes to node $a$. Also, because of their $z$:$x$:$y$ super keys, nodes $c$ and $d$ cannot eliminate their children (nodes $e$, $f$, $g$ and $h$) from recursive search for node $a$'s predecessor node.

Nodes $e$, $f$, $g$ and $h$ have $x$:$y$:$z$ super keys, so their $x$:$y$:$z$ sorted orders relative to node $a$ are known. In that case, selection of a potential predecessor node is similar to selection of the predecessor node in a 1-$d$ tree. Specifically, if any of nodes $e$, $f$, $g$ and $h$ have a \lstinline{null} greater-than child pointer, those nodes ($f$ and $h$) are potential predecessor nodes. But if any of nodes $e$, $f$, $g$ and $h$ do not have a \lstinline{null} greater-than child pointer, those nodes ($e$ and $g$) cannot be potential predecessor nodes; hence, their greater-than child pointers must be followed recursively to inspect potential predecessor nodes $i$ and $l$ respectively. In neither case is it necessary to follow recursively the less-than child pointers of nodes $f$ and $g$ to inspect nodes $j$ and $k$ respectively, because no node in these less-than branches can have the largest $x$:$y$:$z$ super key.

In view of the exclusion of less-than branches, finding 1-child node $a$'s immediate predecessor node requires inspecting $ n^{1-1/k} $ nodes of the $n$-node subtree rooted at node $b$, assuming that subtree to be perfectly balanced with no empty branches \cite{Samet}. And if 1-child node $a$ has only a greater-than child instead of a less-than child, finding node $a$'s immediate successor node requires searching the subtree rooted at that greater-than child to find the node that has the smallest $x$:$y$:$z$ super key.

When the immediate predecessor or immediate successor node to the 1-child node has been found in the manner described above, that predecessor or successor node's \lstinline{tuple} field is copied to the 1-child node's \lstinline{tuple} field, and then the predecessor or successor node is deleted from the tree. But that predecessor or successor node is not necessarily a leaf node, and if not, it must be deleted recursively as one of the three cases itemized above \cite{Rocca}, \cite{Samet}. Recursive deletion converges to a leaf node that is excised from the tree. Then the recursion unwinds, with computation of the height and potential rebalancing at each node along the return path.

The third case is deletion of a node that has two children. This case differs from the 1-child case only in that a deleted 2-child node may be replaced by either its immediate predecessor or its immediate successor node. The choice between the predecessor or successor is arbitrary. However, the optimum replacement node might be selected from the 2-child node's child subtree that has the greater height, in an attempt to improve balance and thereby minimize rebalancing the subtree rooted at the 2-child node \cite{Foster1965}.

Recursive deletion of a non-leaf replacement node always converges to a leaf node that  is excised from the tree. Then the recursion unwinds while computing the height and potentially rebalancing at each node along the return path to the root of the tree.

For the second and third cases, it is possible to curtail recursive deletion before convergence to a leaf node when a subtree contains 3 nodes or fewer. For example, non-leaf node $\left( 8,3,2 \right)$ may be deleted from the tree depicted in Figure \ref{fig:balanced_tree} by removing that node and replacing it with a subtree built from nodes $\left( 8,1,5 \right)$ and $\left( 9,2,1 \right)$. This rebuild requires only comparison of $z$:$x$:$y$ super key 5:8:1 to $z$:$x$:$y$ super key 1:9:2.

\subsection{Benchmark Methodology}

To assess the performance of the dynamic \emph{k}-d tree, benchmarks were executed on a Hewlett-Packard Pro Mini 400 G9 with 2x32GB DDR5-4800 RAM and a 14th-generation Intel Raptor Lake CPU (i7 14700T with 8 performance cores, 5.2GHz performance core maximum frequency, 78.6GB/s maximum memory bandwidth, 80KB per-core L1 and 2MB per-core L2 caches, and a 33MB L3 cache shared by all cores).

A \emph{k}-d tree benchmark measured algorithm performance for trees that contained $n$ nodes, for values of $n$ in the range $ \left[ 1,003,201; 4,523,071 \right] $ that map to equally spaced values of $n \log_2 \left( n \right)$ in the range $\left[ 20,000,000; 100,000,000 \right]$. Each node of the tree stored a $k$-dimensional tuple of 64-bit integers. The integers were equally spaced across the maximum 64-bit integer range $r$, so the spacing was $r/n$. The integers were randomly shuffled via the \lstinline{std::mt19937_64} Mersenne Twister pseudo-random number generator \cite{Matsumoto} and copied to the first of the $k$ dimensions, then randomly shuffled again and copied to the second of the $k$ dimensions, et cetera. So that all benchmarks randomly shuffled the integers in an identical sequence, each benchmark initialized \lstinline{std::mt19937_64} to \lstinline{std::mt19937_64::default_seed}.

The \emph{k}-d tree benchmark was implemented in C++, compiled via Gnu g++ 13.2.0 with  \lstinline{-std=c++20}, \lstinline{-O3}, \lstinline{-pthread -D NLOGN} options to specify red-black balance and a $O \left[ n \log \left( n \right) \right]$ rebuilding algorithm, and executed under Ubuntu 24.04.1 LTS via a single thread mapped to a single performance core via the Ubuntu \lstinline{taskset} command.

Each benchmark measured the execution times for insertion of all tuples into the tree, verification of correct ordering of the tree, search for all tuples in the tree, and deletion of all tuples from the tree via the \lstinline{std::chrono::steady_clock::now()} function. To provide a comparison, the execution time was measured for construction of a static \emph{k}-d tree via a $O \left[ n \log \left( n \right) \right]$ algorithm \cite{Brown2015}. Each benchmark was repeated 100 times and the mean values and standard deviations of the execution times were calculated. The standard deviations were less than 1\% of the mean values.

To assess the performance of the dynamic \emph{k}-d tree for worst-case data, a set of tuples was created by constructing a static \emph{k}-d tree, and then sweeping that tree in sorted order to obtain a set of tuples to insert into and delete from a dynamic \emph{k}-d tree.

\subsection{Benchmark Results}

\begin{figure}[h]
\centering
\centerline{\includegraphics*[trim = {1.00in, 3.47in, 1.37In, 1.52In}, clip, width=\columnwidth]{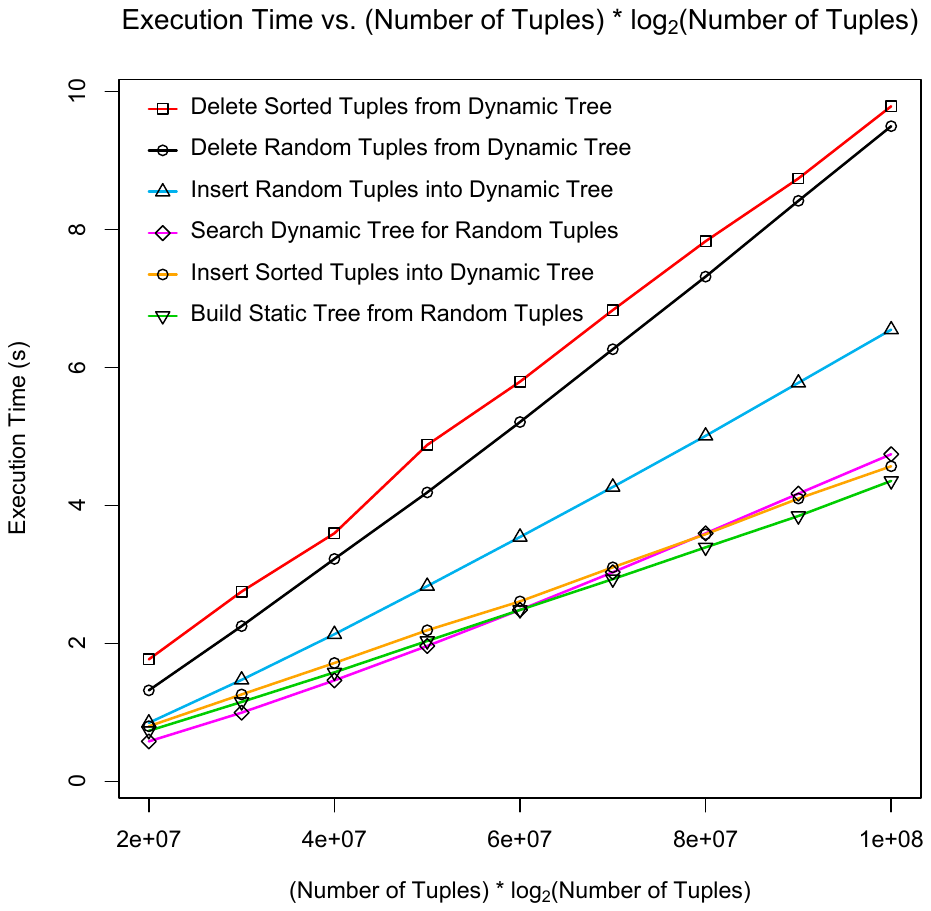}}
\caption{Execution times for insertion into, deletion from, and search of a dynamic \emph{k}-d tree}
\label{fig:execution_times}
\end{figure}

Figure \ref{fig:execution_times} plots execution times (in seconds) for single-threaded insertion into, deletion from, and search of a dynamic \emph{k}-d tree, plotted versus $n \log_2 \left( n \right)$ where $n$ represents the number of tuples in the tree. Execution times are included for random and sorted (worst-case) tuples, as well as for single-threaded construction of a static \emph{k}-d tree.

In Figure \ref{fig:execution_times}, the execution-time plots for insertion, deletion, and search of random and sorted tuples in a dynamic \emph{k}-d tree, and for creation of a static \emph{k}-d tree from random tuples, exhibit $O \left[ n \log \left( n \right) \right]$ computational complexity amortized over $n$ tuples, as expected for a balanced binary tree. The execution time for insertion of random tuples is about 1.5 times longer than the execution time for creating a static \emph{k}-d tree en masse from random tuples. This disparity represents the difference between rebalancing piecemeal for a dynamic tree, versus balancing en masse for a static tree.

Figure \ref{fig:execution_times} reveals that the execution time for insertion of random tuples exceeds the execution time for insertion of sorted tuples. This difference in execution times is likely due to cache memory effects. For insertion of sorted 3-dimensional $\left(x,y,z\right)$ tuples, two consecutive tuples descend the nascent tree via a similar path. Hence, the nodes that the first tuple causes to be loaded into cache during its descent of the tree likely remain resident in cache and are likely accessed by the second tuple during its descent of the tree. In contrast, for insertion of random tuples, it is unlikely that two consecutive tuples descend the tree via a similar path. Hence, the nodes loaded into cache during the first tuple's descent of the tree are unlikely to be accessed by the second tuple. A similar disparity in execution times between sorted and random keys has been observed for the AVL and red-black trees \cite{Brown2025}.

Figure \ref{fig:execution_times} also reveals that the execution time for deletion of sorted tuples exceeds the execution time for deletion of random tuples. Compared to insertion of sorted tuples, it is less likely for deletion of sorted tuples that two consecutive tuples descend the tree via a similar path. This lower probability of a similar path arises because, during the prior insertion of sorted tuples, the tree was considerably restructured via rebalancing. So it is unlikely that, during deletion, nodes loaded into cache during the first tuple's descent of the tree will be accessed by the second tuple. Hence, it is not expected that deletion of sorted tuples would be faster than deletion of random tuples.

Moreover, Table \ref{table:rebuildsizes} provides a possible reason that deletion of sorted tuples is slower than deletion of random tuples. This table reveals that the largest subtree rebuilt for deletion of sorted tuples is 2 to 6 times larger than the largest subtree rebuilt for deletion of random tuples. Rebuilding a larger subtree requires more time than rebuilding a smaller subtree.

\newcolumntype{T}{>{\hsize=2\hsize}X}

\begin{table}[htb]

\begin{tcolorbox}[tab2,tabularx={T||Y|Y|Y|Y|Y|Y|Y|Y|Y|YYYY}]
$n \log_2 \left(n\right)$ & 2e7 & 3e7 & 4e7 & 5e7 & 6e7 & 7e7 & 8e7 & 9e7 & 1e8 \\\hline
$n / 1,000$ &  1,003 & 1,465 & 1,917 & 2,362 & 2,801 & 3,237 & 3,669 & 4,097 & 4,523  \\\hline\hline
Sorted \newline rebuild \newline size/1,000 & 1,003 & 1,465 & 1,917 & 2,361 & 1,618 & 3,234 & 3,668 & 2,985 & 4,523 \\\hline
Random \newline rebuild \newline size/1,000 & 461 & 723 & 728 & 633 & 505 & 615 & 647 & 566 & 820 \\\hline
\end{tcolorbox}

\caption{\label{table:rebuildsizes}
Largest subtree sizes (in units of kilo-nodes)}

\end{table}

\newpage

\begin{figure}[h]
\centering
\centerline{\includegraphics*[trim = {1.00in, 3.47in, 1.37In, 1.52In}, clip, width=\columnwidth]{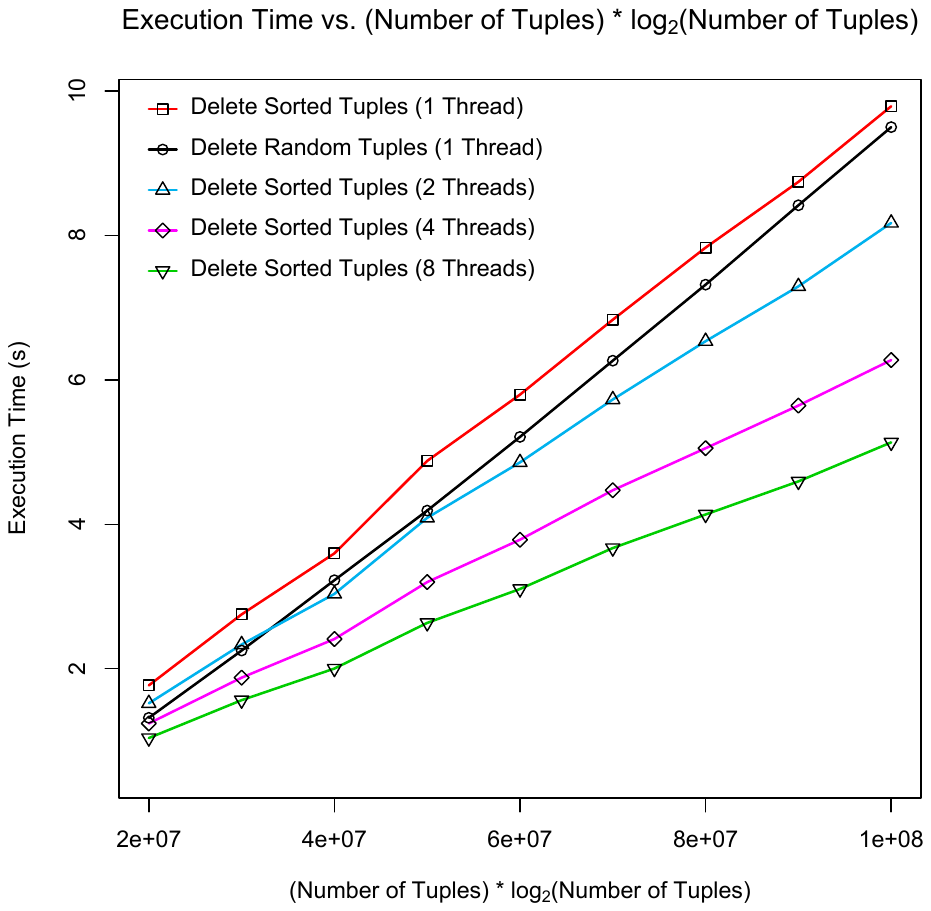}}
\caption{Execution times for deletion of sorted (worst-case) tuples from a dynamic \emph{k}-d tree}
\label{fig:multithread_times}
\end{figure}

The prolonged subtree-rebuilding times for sorted tuples may be mitigated by rebuilding via multiple threads, as supported by the $O \left[ n \log \left( n \right) \right]$ algorithm. Figure \ref{fig:multithread_times} plots execution times (in seconds) for deletion of sorted tuples from a dynamic \emph{k}-d tree via 1 to 8 threads mapped to up to 8 performance cores specified by the Ubuntu \lstinline{taskset} command. The execution time for single-threaded deletion of random tuples is included from Figure \ref{fig:execution_times} for comparison. The plots reveal that 2 to 8 threads shorten the execution time significantly relative to 1 thread for deletion of sorted tuples.

Multi-threaded execution of the $O \left[ n \log \left( n \right) \right]$ algorithm should be used judiciously. Small subtrees are rebuilt frequently, and the time required to spawn child threads for a small subtree outweighs the benefit of those threads to rebuilding that subtree. The execution-time data plotted in Figure \ref{fig:multithread_times} were obtained using multiple threads for only subtrees that contained more than 65,536 nodes.

\newpage

\subsection{Choosing the Replacement Node for a Deleted Two-child Node}

Section \ref{sec:Deletion} proposes that a deleted 2-child node may be replaced by either its immediate predecessor node or its immediate successor node, and that the replacement node may be obtained from the higher of the deleted node's child subtrees. Table \ref{table:replacement} reports execution times (in seconds) for deletion of random tuples from a \emph{k}-d tree, wherein the replacement node for a deleted 2-child node is obtained either from the higher child subtree or always from the greater-than subtree (i.e., the immediate successor node). This table reveals that neither node replacement strategy is consistently faster than the other, and the difference between deletion execution times is 1\% or less.

\newcolumntype{U}{>{\hsize=2\hsize}X}

\begin{table}[htb]

\begin{tcolorbox}[tab2,tabularx={U||Y|Y|Y|Y|Y|Y|Y|Y|Y|YYYY}]
$n \log_2 \left(n\right)$ & 2e7 & 3e7 & 4e7 & 5e7 & 6e7 & 7e7 & 8e7 & 9e7 & 1e8 \\\hline
$n / 1,000$ &  1,003 & 1,465 & 1,917 & 2,362 & 2,801 & 3,237 & 3,669 & 4,097 & 4,523  \\\hline\hline
Higher \newline subtree & 1.32 & 2.25 & 3.23 & 4.19 & 5.21 & 6.27 & 7.32 & 8.42 & 9.50 \\\hline
Successor \newline node & 1.33 & 2.26 & 3.23 & 4.21 & 5.21 & 6.26 & 7.32 & 8.40 & 9.50 \\\hline
\end{tcolorbox}

\caption{\label{table:replacement}
Deletion execution times (s) for two strategies to choose a replacement node}

\end{table}

\subsection{AVL Versus Red-black Balance}

Section \ref{sec:Balance} of this article explains that the balance of a \emph{k}-d tree may be established either via the red-black criterion, which requires that the heights of the two child subtrees of a node differ by at most a factor of two; or via the AVL criterion, which requires that the heights of the two child subtrees of a node differ by 1. In fact, it is possible to define the AVL criterion less stringently; for example,  to allow the subtrees to differ by an integer in the range $\left[1,4\right]$ instead of by 1 \cite{Foster1973}.

\begin{figure}[h]
\centering
\centerline{\includegraphics*[trim = {1.00in, 3.47in, 1.37In, 1.52In}, clip, width=\columnwidth]{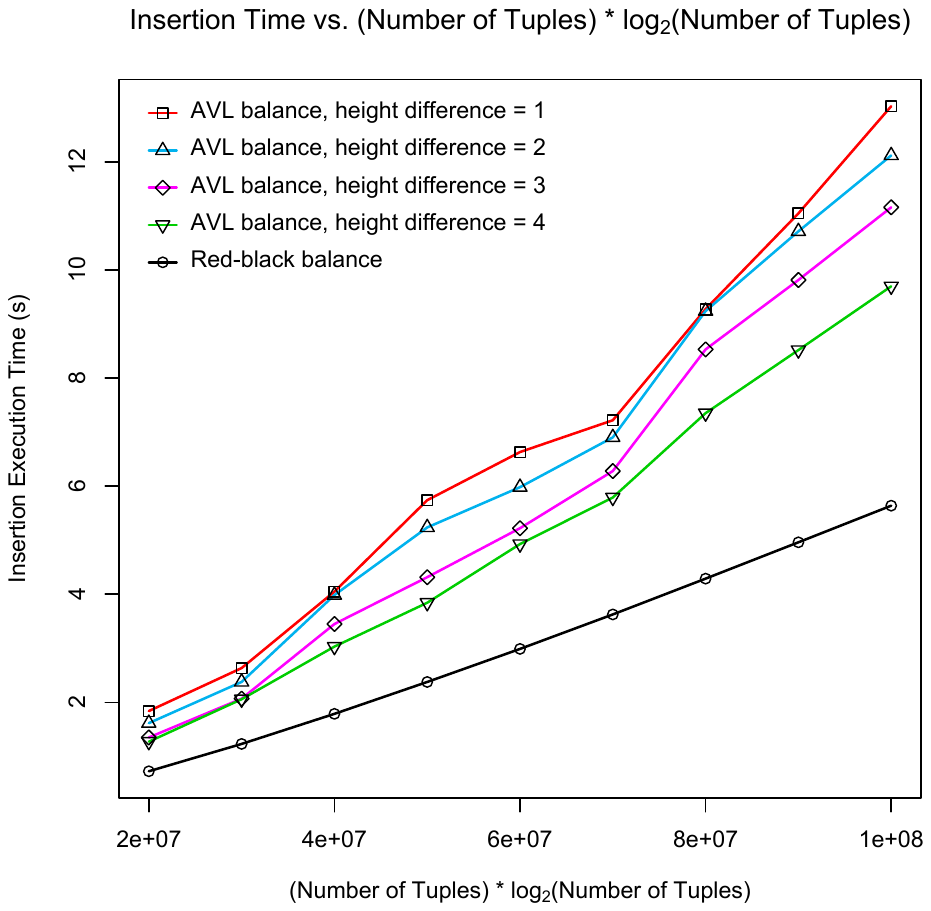}}
\caption{Execution times for insertion of random tuples into a dynamic \emph{k}-d tree}
\label{fig:insertion}
\end{figure}

Figure \ref{fig:insertion} plots execution times (in seconds) for insertion of random tuples into \emph{k}-d trees that were balanced according to AVL criteria for height differences of 1 to 4,  or according to the red-black criterion. This figure reveals that the insertion execution times increase for decreasing AVL balance height differences, and that the insertion execution time for red-black balance is less than those for AVL balance. The plot for red-black balance is smooth, whereas the plots for AVL balance exhibit variation that increases with decreasing height differences. This variation is likely due to the fact that AVL balance requires rebuilding larger subtrees than red-black balance.

Table \ref{table:AVLredblackinsertion} reports the size (in units of kilo-nodes) of the largest subtree rebuilt after insertion of random tuples, for trees that contain $n$ tuples. The substantially larger subtrees for AVL balance introduce larger quanta of execution time; hence, the plots for AVL balance exhibit larger variation than the plot for red-black balance.

\newpage

\newcolumntype{V}{>{\hsize=2\hsize}X}

\begin{table}[htb]

\begin{tcolorbox}[tab2,tabularx={V||Y|Y|Y|Y|Y|Y|Y|Y|Y|YYYY}]
$n \log_2 \left(n\right)$ & 2e7 & 3e7 & 4e7 & 5e7 & 6e7 & 7e7 & 8e7 & 9e7 & 1e8 \\\hline
$n / 1,000$ & 1,003 & 1,465 & 1,917 & 2,362 & 2,801 & 3,237 & 3,669 & 4,097 & 4,523 \\\hline\hline
AVL 1 \newline / 1,000 & 1,002 & 1,045 & 1,907 & 2,095 & 2,081 & 2,066 & 3,662 & 4,062 & 4,102 \\\hline
AVL 4 \newline / 1,000 & 999 & 1,448 & 1,805 & 2,087 & 2,380 & 3,215 & 3,606 & 4,030 & 4,313 \\\hline
Red-black \newline / 1,000 & 0.622 & 0.626 & 0.556 & 0.619 & 0.907 & 0.657 & 0.631 & 0.893 & 1.120 \\\hline
\end{tcolorbox}

\caption{\label{table:AVLredblackinsertion}
Largest subtree sizes for AVL and red-black insertion (in units of kilo-nodes)}

\end{table}

\newpage

\begin{figure}[h]
\centering
\centerline{\includegraphics*[trim = {1.00in, 3.47in, 1.37In, 1.52In}, clip, width=\columnwidth]{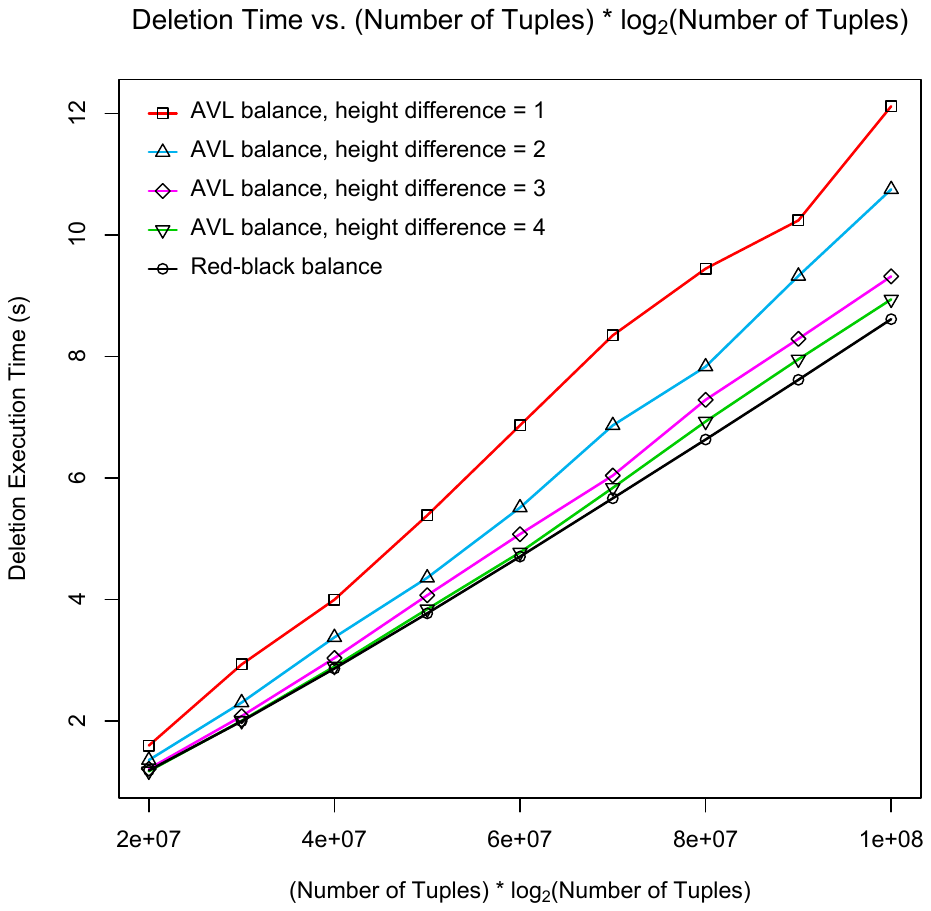}}
\caption{Execution times for deletion of random tuples from a dynamic \emph{k}-d tree}
\label{fig:deletion}
\end{figure}

Figure \ref{fig:deletion} plots execution times (in seconds) for deletion of random tuples from \emph{k}-d trees that were balanced according to AVL criteria for height differences of 1 to 4,  or according to the red-black criterion. This figure reveals that the deletion execution times increase for decreasing AVL balance height differences, and that the deletion execution time for red-black balance is less than those for AVL balance. The plot for red-black balance is smooth, whereas the plots for AVL balance exhibit variation that increases with decreasing height differences. This variation is likely due to the fact that AVL balance requires rebuilding larger subtrees than red-black balance.

Table \ref{table:AVLredblackdeletion} reports the size (in units of kilo-nodes) of the largest subtree rebuilt after deletion of random tuples, for trees that contain $n$ tuples. The larger subtrees for AVL balance introduce larger quanta of execution time; hence, the plots for AVL balance exhibit larger variation than the plot for red-black balance.

\newpage

\begin{table}[htb]

\begin{tcolorbox}[tab2,tabularx={V||Y|Y|Y|Y|Y|Y|Y|Y|Y|YYYY}]
$n \log_2 \left(n\right)$ & 2e7 & 3e7 & 4e7 & 5e7 & 6e7 & 7e7 & 8e7 & 9e7 & 1e8 \\\hline
$n / 1,000$ & 1,003 & 1,465 & 1,917 & 2,362 & 2,801 & 3,237 & 3,669 & 4,097 & 4,523 \\\hline\hline
AVL 1 \newline / 1,000 & 1,001 & 1,025 & 1,904 & 2,057 & 2,073 & 2,089 & 3,662 & 4,080 & 4,136 \\\hline
AVL 4 \newline / 1,000 & 961 & 1,464 & 9.011 & 4.788 & 2,198 & 3,217 & 3,639 & 3,338 & 3,464 \\\hline
Red-black \newline / 1,000 & 0.674 & 0.723 & 0.962 & 0.889 & 0.770 & 1.156 & 1.008 & 1.301 & 1.002 \\\hline
\end{tcolorbox}

\caption{\label{table:AVLredblackdeletion}
Largest subtree sizes for AVL and red-black deletion (in units of kilo-nodes)}

\end{table}

\begin{figure}[h]
\centering
\centerline{\includegraphics*[trim = {1.00in, 3.47in, 1.37In, 1.52In}, clip, width=\columnwidth]{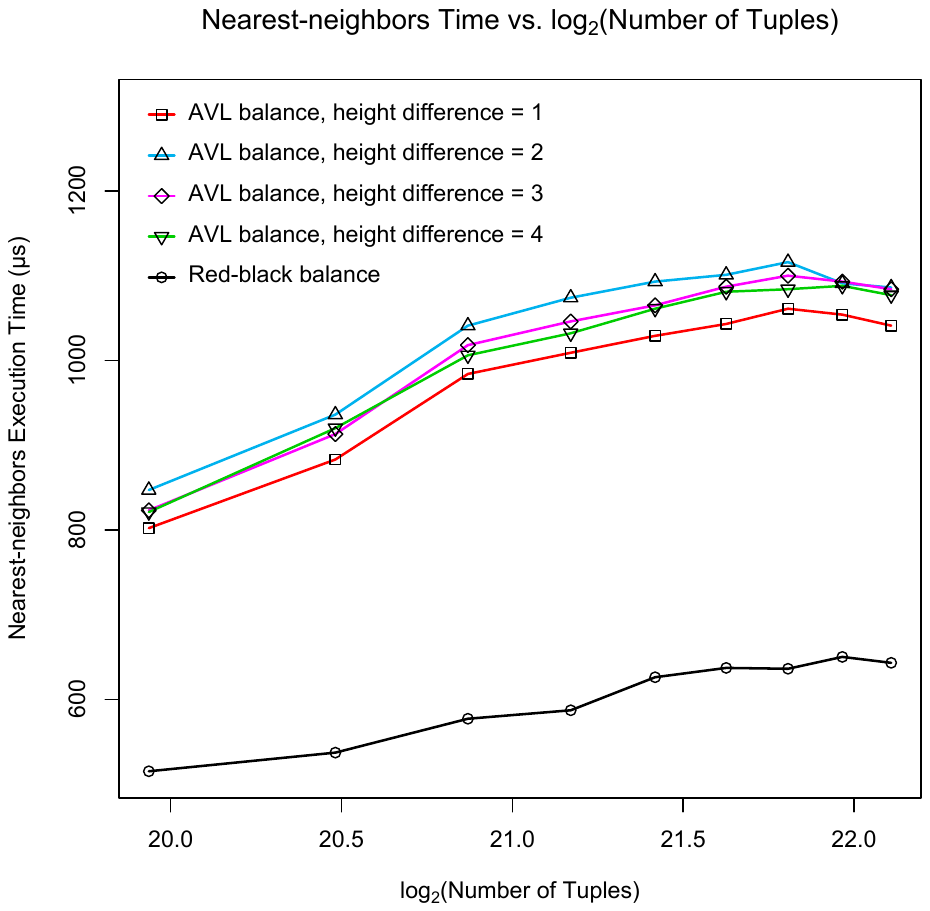}}
\caption{Execution times for nearest-neighbor search of a dynamic \emph{k}-d tree}
\label{fig:nearestneighbors}
\end{figure}

\newpage

Figure \ref{fig:nearestneighbors} plots the execution times (in microseconds) of search for the 1,000 nearest neighbors to tuple $\left( 0,1,2 \right)$ for trees that contain $n$ tuples. Because the $n$ tuples are distributed randomly on an equally spaced 3-dimensional grid, the choice of tuple $\left( 0,1,2 \right)$ is arbitrary, and any other tuple would provide equivalent results. In Figure \ref{fig:nearestneighbors}, the $x$-axis is $\log_2\left(n\right)$ because the worst-case performance of nearest-neighbor search is expected to be $O\left[\log_2\left(n\right)\right]$ \cite{Friedman}. This figure reveals that the execution times for AVL balance exceed the execution time for red-black balance.

\begin{figure}[h]
\centering
\centerline{\includegraphics*[trim = {1.00in, 3.47in, 1.37In, 1.52In}, clip, width=\columnwidth]{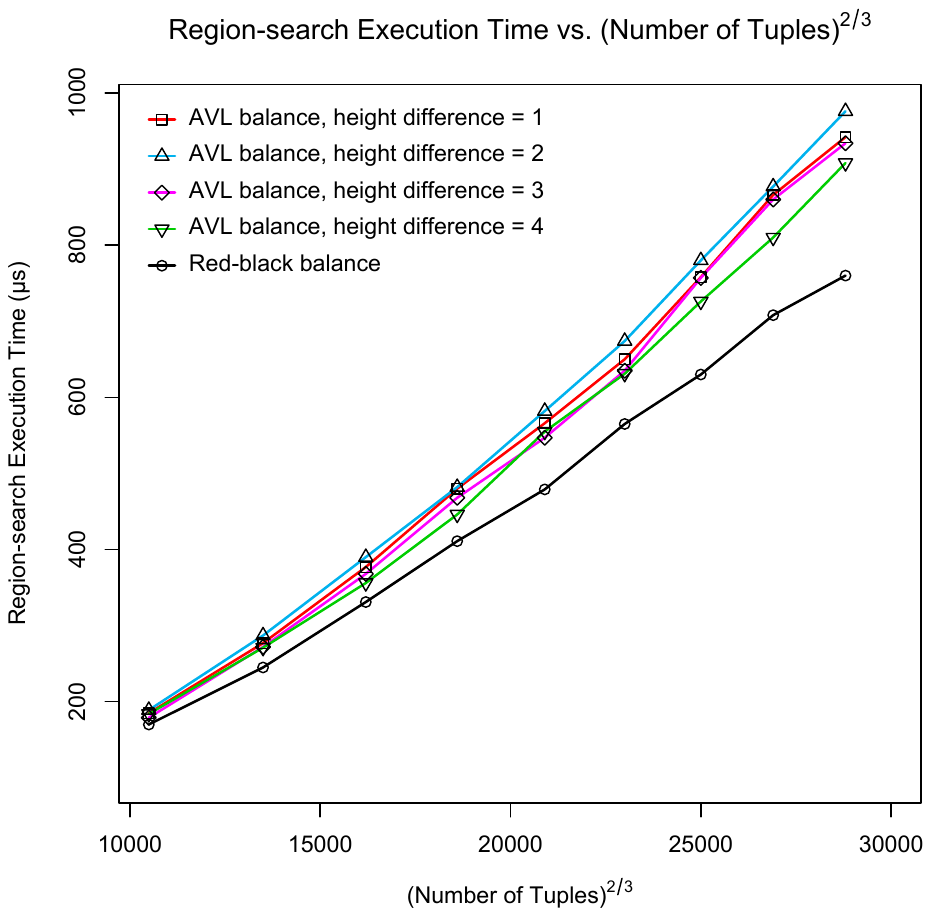}}
\caption{Execution times for region search of a dynamic \emph{k}-d tree}
\label{fig:regionsearch}
\end{figure}

Figure \ref{fig:regionsearch} plots the execution times (in microseconds) of search for all tuples that lie within a cubic region that is centered at tuple $\left( 0,1,2 \right)$ and that occupies $1/1,000^{\mathrm{th}}$ of the 3-dimensional space, for trees that contain $n$ tuples. The $x$-axis in this figure is $n^{2/3}$ because the worst-case performance of region search is expected to be $O\left(n^{1-1/k}\right) = O\left(n^{2/3}\right)$ for $k=3$ \cite{Lee}. This figure reveals that the execution times for AVL balance exceed the execution time for red-black balance.

\newpage

The improved search performance for red-black balance compared to AVL balance, as revealed by Figures \ref{fig:nearestneighbors} and \ref{fig:regionsearch}, appears to contradict the expectation that search of an AVL tree should outperform search of an equivalent red-black tree because the average height of a red-black tree is greater than the average height of the equivalent AVL tree \cite{Wirth}. And although Table {\ref{table:AVLheight} reveals that the heights of \emph{k}-d trees balanced via the red-black criterion exceed the heights of \emph{k}-d trees balanced via the AVL criteria for height differences of 1 to 4, red-black balance nevertheless outperforms AVL balance for nearest-neighbor search and region search. A possible explanation for this discrepancy is that Table \ref{table:AVLheight} reports maximum heights, whereas the expectation of AVL tree outperformance is based on a consideration of average heights.

\newcolumntype{W}{>{\hsize=2\hsize}X}

\begin{table}[htb]

\begin{tcolorbox}[tab2,tabularx={W||Y|Y|Y|Y|Y|Y|Y|Y|Y|YYYY}]
$n \log_2 \left(n\right)$ & 2e7 & 3e7 & 4e7 & 5e7 & 6e7 & 7e7 & 8e7 & 9e7 & 1e8 \\\hline
$n / 1,000$ & 1,003 & 1,465 & 1,917 & 2,362 & 2,801 & 3,237 & 3,669 & 4,097 & 4,523 \\\hline\hline
AVL 1 \newline height & 22 & 23 & 23 & 23 & 23 & 24 & 24 & 24 & 24 \\\hline
AVL 2 \newline height & 22 & 24 & 23 & 24 & 24 & 26 & 26 & 25 & 26 \\\hline
AVL 3 \newline height & 25 & 25 & 26 & 25 & 27 & 27 & 27 & 27 & 26 \\\hline
AVL 4 \newline height & 26 & 27 & 27 & 28 & 28 & 27 & 28 & 27 & 26 \\\hline
Red-black \newline height & 30 & 32 & 32 & 32 & 34 & 33 & 33 & 33 & 34 \\\hline
\end{tcolorbox}

\caption{\label{table:AVLheight}
Tree heights for AVL and red-black balance}

\end{table}

\section{Conclusions}
\label{Conclusions}

A dynamic \emph{k}-d tree confers flexibility to building a \emph{k}-d tree. Instead of building a static, balanced \emph{k}-d tree en masse from all of the \emph{k}-dimensional data, it is possible to insert into or delete from the tree one datum at a time. Alternately, it is possible to build a static, balanced \emph{k}-d tree from the data available at a particular moment, and thereafter modify the tree dynamically via insertion or deletion of data that become available subsequently, one datum at a time.

Rebalancing a dynamic \emph{k}-d tree requires rebuilding subtrees. Although rebalancing via rebuilt subtrees has been proposed previously \cite{Overmars}, the present manuscript appears to be the first to report implementation and performance of a dynamic \emph{k}-d tree rebalanced via the rebuilt-subtrees method.

\newpage

A dynamic \emph{k}-d tree is able to self-balance if it stores a \lstinline{height} field at each node of the tree, where height is defined as the maximum path length from that node to the bottom of the tree. Following recursive insertion or deletion, the height is recomputed at each node along the path from the point of insertion or deletion to the root of the tree as the recursion unwinds. If the balance at any node indicates that the tree has become unbalanced as a result of the insertion or deletion, the subtree rooted at that node is rebuilt to rebalance the tree. Insertion and deletion, each including rebalancing, exhibit $O \left[ n \log \left( n \right) \right] $ computational complexity for a dynamic \emph{k}-d tree.

Balance is determined using either an AVL criterion or a red-black criterion. The red-black criterion confers better performance than the AVL criterion for insertion, deletion, nearest-neighbor search, and region search.

Further research could include: (1) collecting histograms of rebuilt subtrees sizes for insertion into or deletion from a dynamic \emph{k}-d tree to assess whether execution times correlate with rebuilt subtree sizes, (2) comparing the average height to the maximum height at each node of a dynamic \emph{k}-d tree to determine whether these two heights are correlated, and (3) comparing the performance of a single dynamic \emph{k}-d tree to that of multiple static \emph{k}-d trees built via the logarithmic method \cite{Overmars1981}.

\section*{Supplemental Materials}

Included with this manuscript are Java and C++ implementations of $O \left[ n \log \left( n \right) \right]$  algorithms that insert into and delete from a dynamic \emph{k}-d tree or a dynamic \emph{k}-d tree-based key-to-multiple-values map. Also included are $O \left[ n \log \left( n \right) \right]$  and $O \left[ kn \log \left( n \right) \right]$ algorithms that build a static \emph{k}-d tree or a static \emph{k}-d tree-based key-to-multiple-values map and that rebalance a subtree of the dynamic tree or map.

The dynamic \emph{k}-d tree is implemented as the \lstinline{KdTreeDynamic} class that derives from (in C++), or extends (in Java), the \lstinline{KdTree} class that implements the $O \left[ n \log \left( n \right) \right]$  and $O \left[ kn \log \left( n \right) \right]$ algorithms that build a static \emph{k}-d tree. Hence, all other algorithms implemented by the \lstinline{KdTree} class, such as  region search, nearest-neighbor search and reverse-nearest-neighbor search \cite{Korn}, are accessible by the \lstinline{KdTreeDynamic} class. In addition, it is possible to build a static \emph{k}-d tree via an $O \left[ n \log \left( n \right) \right]$  or $O \left[ kn \log \left( n \right) \right]$ algorithm, and then provide that static tree to the \lstinline{KdTreeDynamic} constructor so that the static tree may thereafter be updated dynamically via insertion or deletion of individual tuples.

The Java and C++ implementations are also available at the following URL.

\smallskip
\href{https://github.com/RussellABrown/kd-tree}{https://github.com/RussellABrown/kd-tree}

\section*{Acknowledgements}

The author thanks John A. Robinson for helpful discussions.

\newpage

\section*{Author Contact Information}

\href{https://www.linkedin.com/in/russellabrown/}{https://www.linkedin.com/in/russellabrown/}


\section*{References}

\small
\bibliographystyle{jcgt}
\bibliography{paper}

\end{document}